\documentclass[aps,prd,nofootinbib,showpacs,superscriptaddress,preprint]{revtex4-1}
\pdfoutput=1
\usepackage[utf8]{inputenc}
\usepackage{amsmath,amssymb}
\usepackage{graphicx}
\usepackage{multirow}
\usepackage[usenames,dvipsnames]{color}
\usepackage{babel}
\usepackage{float}
\usepackage[colorlinks,citecolor=blue]{hyperref}
\usepackage[compat=1.0.0]{tikz-feynman}
\usepackage{color}
\usepackage{subfigure}
\DeclareUnicodeCharacter{2212}{-}
\newcommand{\orcid}[1]{\href{https://orcid.org/#1}{\resizebox{10px}{!}{\includegraphics{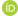}}}}
\begin{document}
\title{  Gamma Ray Bursters and Black Holes in  Gravity's Rainbow}

\author{Moh Vaseem Akram\orcid{0000-0001-7606-6717}}
\email{moh158947@st.jmi.ac.in}
\affiliation{Department of Physics, Jamia Millia Islamia, New Delhi, India}

\author{Imtiyaz Ahmad Bhat\orcid{0000-0002-2695-9709}}
\email{imtiyaz@ctp-jamia.res.in}
\affiliation{Center for Theoretical Physics, Jamia Millia Islamia, New Delhi, India}
\affiliation{Department of Physics, Central University of Kashmir, Ganderbal 191201 India}
\author{Anver Aziz\orcid{0000-0003-1762-7558}} 
\email{aaziz@jmi.ac.in}
\affiliation{Department of Physics, Jamia Millia Islamia, New Delhi, India}
\begin{abstract}
In this letter, we analyze the modification to the thermodynamics of a Schwarzschild  black hole and a Kerr black hole due to gravity's rainbow. The metric for these black holes will be made energy dependent. This will be done by using    rainbow functions   motivated from    the hard spectra from gamma-ray bursters at cosmological distances.
This modification of the metric by these rainbow functions will in turn modify the temperature and entropy of these black holes. We will also discuss how this effects the formation of virtual black holes. 
\end{abstract}
\maketitle

It is known that the  string theory can be investigated using the formalism of a  two dimensional conformal field theory. In this approach, the  metric in the   target space can be viewed as   
a matrix of coupling constants for the conformal field theory. Now it will be possible for these coupling constants to flow because of the   renormalization group flow associated with the quantum field theory describing the world-sheet of a string \cite{rosten2012fundamentals,warner2000renormalization}. This flow would make the target space metric depend on the scale at which it is being probed. However, the scale at which the metric is probed will depend on the energy of the probe, and so it is expected that the spacetime metric will become energy dependent. A theory, where the metric depends on the energy of the probe is called     gravity's rainbow \cite{de2018gravitational,ali2015proposal,assanioussi2017rainbow,hendi2015black}. Such modification to gravity has also been proposed due to  loop quantum gravity \cite{amelino1997distance,amelino2013quantum}, discrete spacetime \cite{hooft1996quantization}, string field theory \cite{hooft1996quantization},
spacetime foam \cite{kostelecky1989spontaneous}, spin-networks \cite{gambini1999nonstandard},  non-commutative geometry \cite{carroll2001noncommutative,faizal2013noncommutative}, and ghost condensation \cite{faizal2011spontaneous}.

In gravity's rainbow  corrections  to the thermodynamics of black holes occurs at very small scales \cite{ali2014remnants,ali2015remnant}. This modified thermodynamics of black holes agrees with the usual thermodynamics, for large black holes. However, it differs considerably  for small black holes.  This difference in the behavior of thermodynamics can have important consequences for the detection of mini black holes at the  LHC \cite{ali2015absence}. The modification to the thermodynamics of black branes has also been studied using gravity's rainbow \cite{ashour2016branes}. The modification to the thermodynamics in Vaidya spacetime due to gravity's rainbow has also been investigated \cite{heydarzade2017vaidya, Rudra:2016alu}. 
The modification to the thermodynamics of Yang-Mills black hole from gravity's rainbow has also been studied \cite{Aounallah:2020yjo}. It has been observed that these rainbow functions also modify the black holes in higher curvature gravity in a non-trivial way \cite{Hendi:2017pld, Hendi:2016njy, Hendi:2015hja}.  The modification to a dilatonic   black hole in gravity's rainbow has also been studied \cite{Hendi:2015cra}. 

Thus, various different black holes have been studied due to gravity's rainbow. However, most of the discussion has been limited to a rainbow function motivated from loop quantum gravity  \cite{Amelino-Camelia:2008aez}. Here, we will use an alternative rainbow function, which is motivated from cosmological data. We will use the   rainbow function    motivated by using  the hard spectra from gamma-ray bursters at cosmological distances \cite{amelino1998tests}. This has been done by observing that  the fine-scale time structure and hard spectra of gamma-ray bursters  emissions are very sensitive to the possible dispersion of electromagnetic waves, and can be used to estimate their modification, which in turn can be absorbed in rainbow functions. We will observe how this changes the last stage of evaporation of black holes.  It has been proposed that  the virtual black holes occur at the stages of evaporation of black holes \cite{hawking1996virtual,hawking1997loss,faizal2012some,ohkuwa2017virtual}. These virtual black holes form like virtual particles due to quantum fluctuations. However, they also form due to the last stage of evaporation of black holes. Now if the last stage of evaporation of black holes is modified then the formation of virtual black holes will also be modified. In this paper, we will analyze the formation of virtual black holes in gravity's rainbow. We would like to point out that it has been suggested that black hole information paradox might get resolved due to gravity's rainbow \cite{Ali:2015iba, Ali:2014cpa}. Now we propose that this will happen due to the formation of virtual black holes in gravity's rainbow.

Now we will analyze the modification of thermodynamics for Schwarzschild black hole due to  gravity's rainbow. 
The temperature of such a  black hole that can be obtain from the metric expressed  in the form 
\begin{align}
    ds^2=-A(r ){dt}^2+\frac{1}{B(r)}{dr}^2+h_{{ij}}{dx}^i{dx}^j
\end{align}
It is  given by\cite{angheben2005hawking},
\begin{align}
    T_0=\frac{1}{4\pi }\sqrt{A_{,r}\big(r_+\big) B_{,r}\big(r_+\big)}
\end{align}
where $r_+$ is given by the largest radius at which $B(r)=0$.
We will analyze   Schwarzschild black hole in the presence of  gravity's rainbow. We will use the rainbow function motivated using    the hard spectra from gamma-ray bursters at cosmological distances \cite{amelino1998tests} 
\begin{align}
    f(E)=\frac{e^{\alpha \big(\frac{E}{E_p}\big)^n}-1}{\alpha \big(\frac{E}{E_p}\big)^n} \hspace{3cm} g(E)=1
\end{align}
The temperature of a black hole in  gravity's rainbow is modified by 
\begin{align}
    T=T_0\frac{g(E)}{f(E)}
\end{align}
For Schwarzschild black hole $r_+=2M$, and so  $T_0$ for Schwarzschild black hole is given by \cite{ali2014black}.
\begin{align}
    T_0=\frac{1}{8\pi  M}
\end{align}
According to \cite{ali2014remnants,ali2015remnant}, the uncertainty principle $\Delta p \geq \frac{1}{\Delta x}$ can be translated to lower bound on the energy of the particle emitted in Hawking radiation and the value of the uncertainty in position can be taken to be the event horizon radius as
\begin{align}
  E \geq \frac{1}{\Delta x}\cong \frac{1}{r_+}
  \end{align}
Substituting values of $E$ and $M$ in terms of area $A$ the modified temperature of  the Schwarzschild black hole becomes 
\begin{align}
   T=\frac{\alpha }{8\pi  M}\bigg(\frac{1}{r_+E_p}\bigg){}^n\frac{1}{e^{\alpha \big(\frac{1}{r_+E_p}\big){}^n}-1}
\end{align}
\begin{figure}[H]
    \centering
    \includegraphics[scale=0.7]{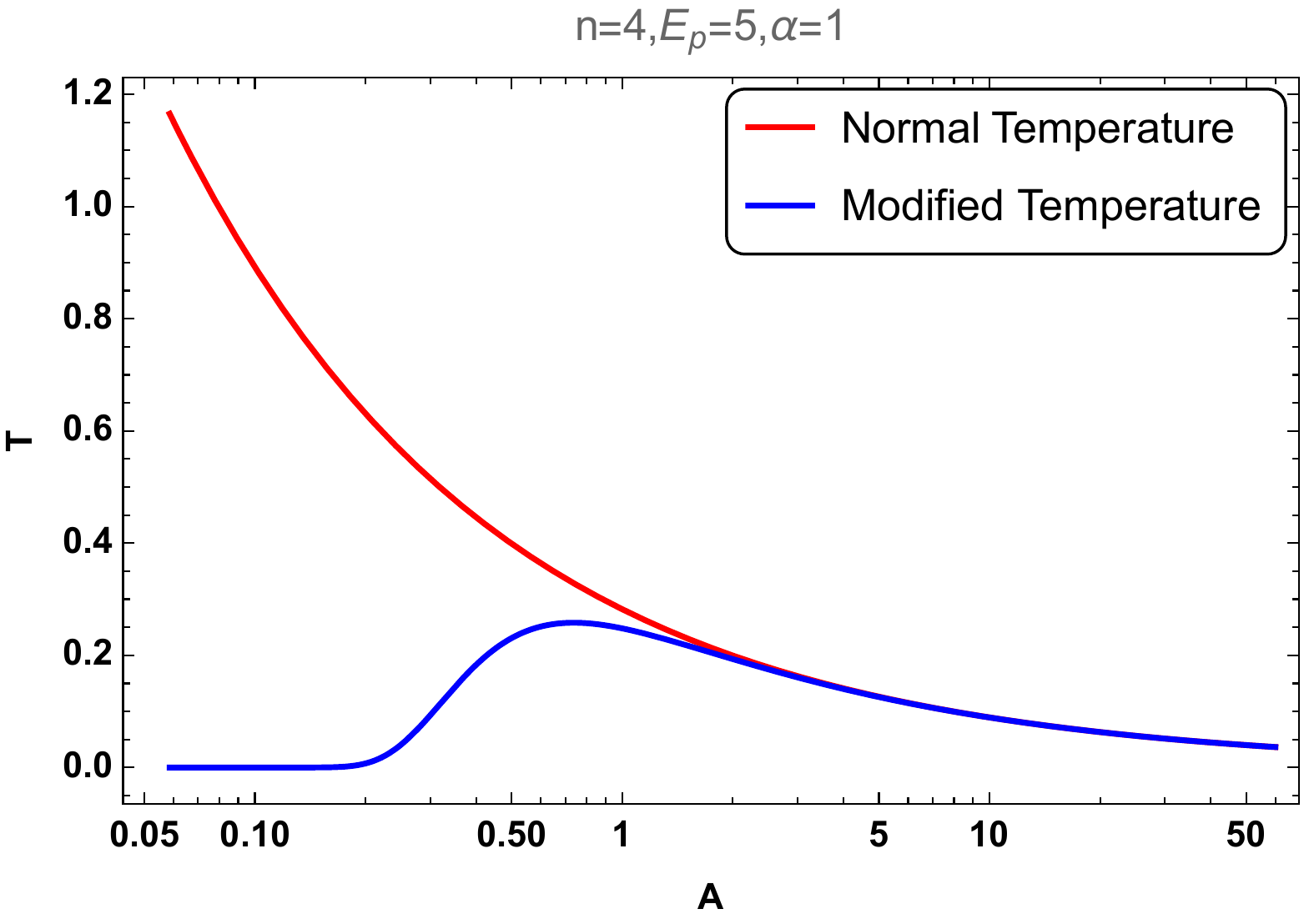}
    \caption{Comparison of normal temperature and modified temperature with area of Schwarzschild black hole}
    \label{fig:scTA}
\end{figure}
The normal and modified temperature of Schwarzschild black hole is plotted in figure \eqref{fig:scTA}.
As can be seen from above equation that the temperature goes to zero at some finite value of surface area. That means the black hole stops radiating and remnant forms. To calculate the entropy, we use the first law of thermodynamics,
\begin{align}\label{ent}
    dS=\frac{dM}{T}
\end{align}
For Schwarschild black hole, $A=4\pi r_+^2$ and $r_+=2M$.  Using above equation \eqref{ent}, we find the expression for entropy in   gravity's rainbow,
\begin{align}
    S=2^{n-4} \alpha  E_p \pi ^{(n/2-1)}\int \frac{\big(\frac{1}{E_p}\sqrt{\frac{1}{A}}\big){}^{n+1}}{\sqrt{A} \big(e^{\alpha \big(\frac{2}{E_p}\sqrt{\frac{\pi
}{A}}\big){}^n }-1\big)} \, dA
\end{align}
\begin{figure}[H]
    \centering
    \includegraphics[scale=0.7]{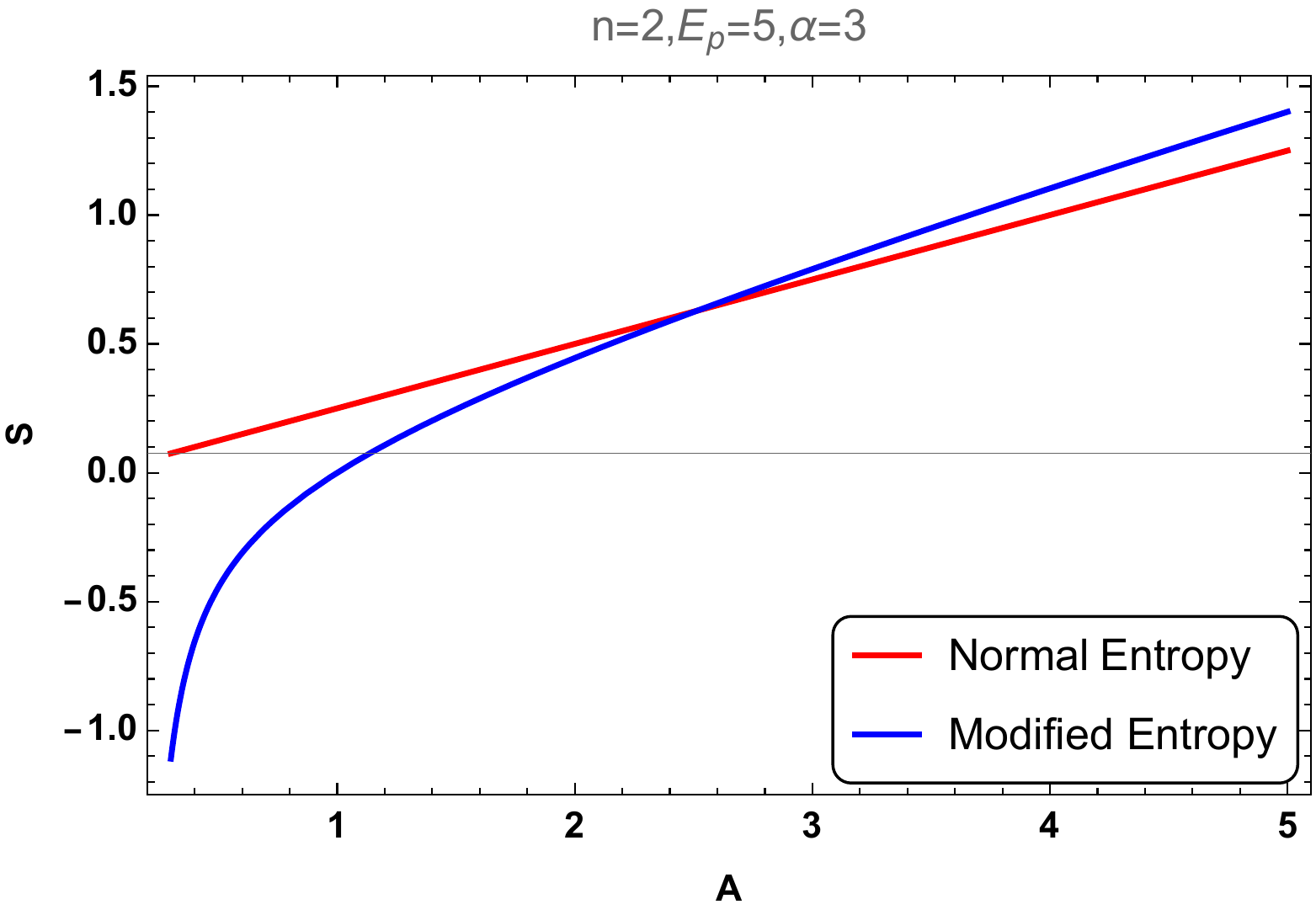}
    \caption{Comparison of normal and modified entropy $S$ versus area of Schwarzschild black hole }
    \label{fig:scSA}
\end{figure}

\begin{figure}[H]
    \centering
    \includegraphics[scale=0.7]{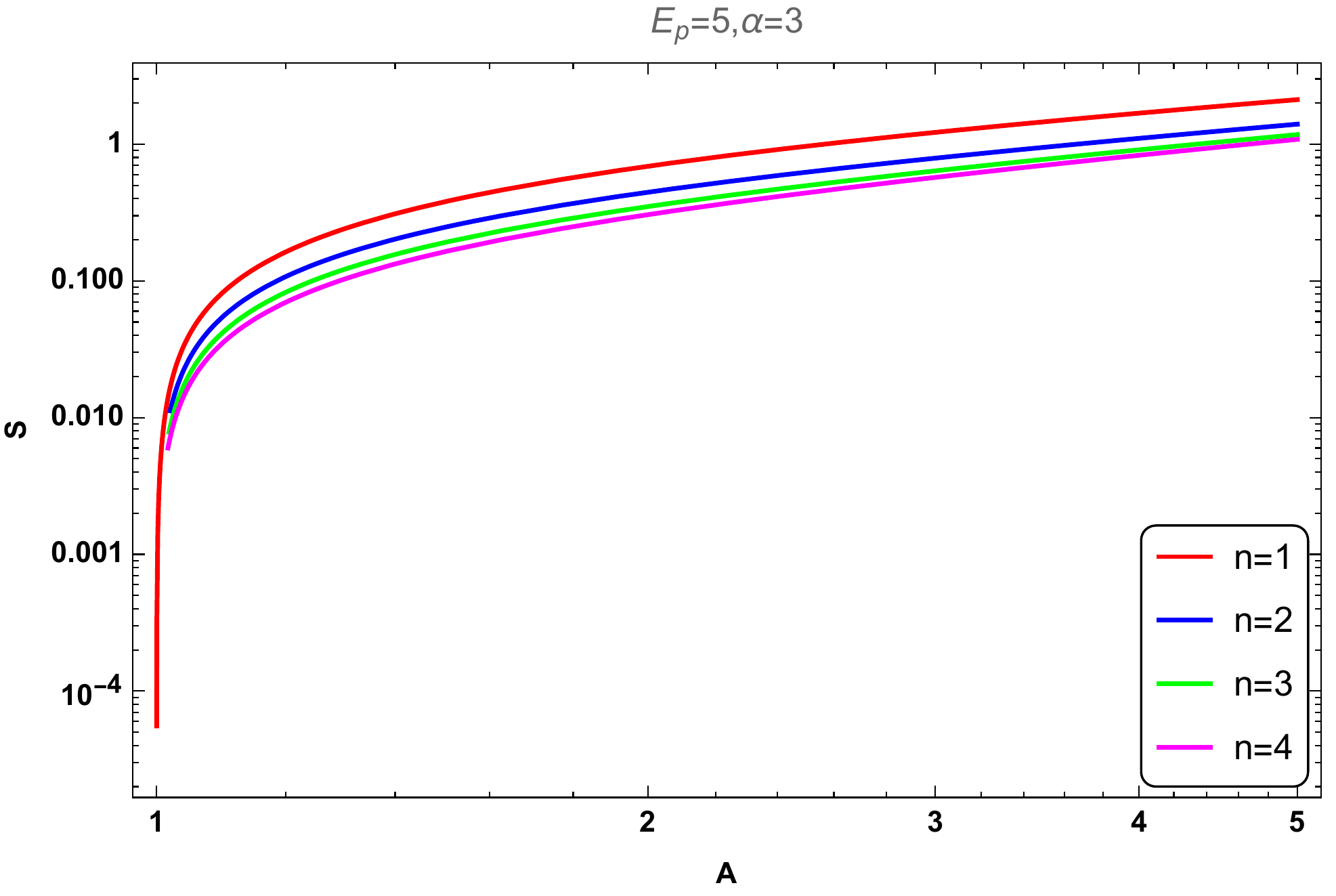}
    \caption{Behaviour of Entropy with $n=1,2,3,4$ for Schwarzschild black hole in gravity's rainbow}
    \label{fig:scSAR}
\end{figure}

The thermodynamic stability of black holes is determined by the heat capacity $C_J$ at constant angular momentum $J$, which is given by,
\begin{align}\label{cj}
    C_J=T\bigg(\frac{\partial S}{\partial  T}\bigg)_J
\end{align}
\begin{align}
    C_J=\frac{ A^{n+1} \big(e^{2^n A^{-n/2} {E_p}^{-n} \pi ^{n/2} \alpha }-1\big)^2 E_p{}^{2 n} \pi ^{-n/2}}{2^{n+1}\alpha  \big(-A^{n/2}
    \big(e^{2^n A^{-n/2} {E_p}^{-n} \pi ^{n/2} \alpha }-1\big) E_p{}^n (n+1)+2^n e^{2^n A^{-n/2} E_p^{-n} \pi ^{n/2} \alpha } n \pi ^{n/2} \alpha
\big)}
\end{align}

\begin{figure}[H]
    \centering
    \includegraphics[scale=0.7]{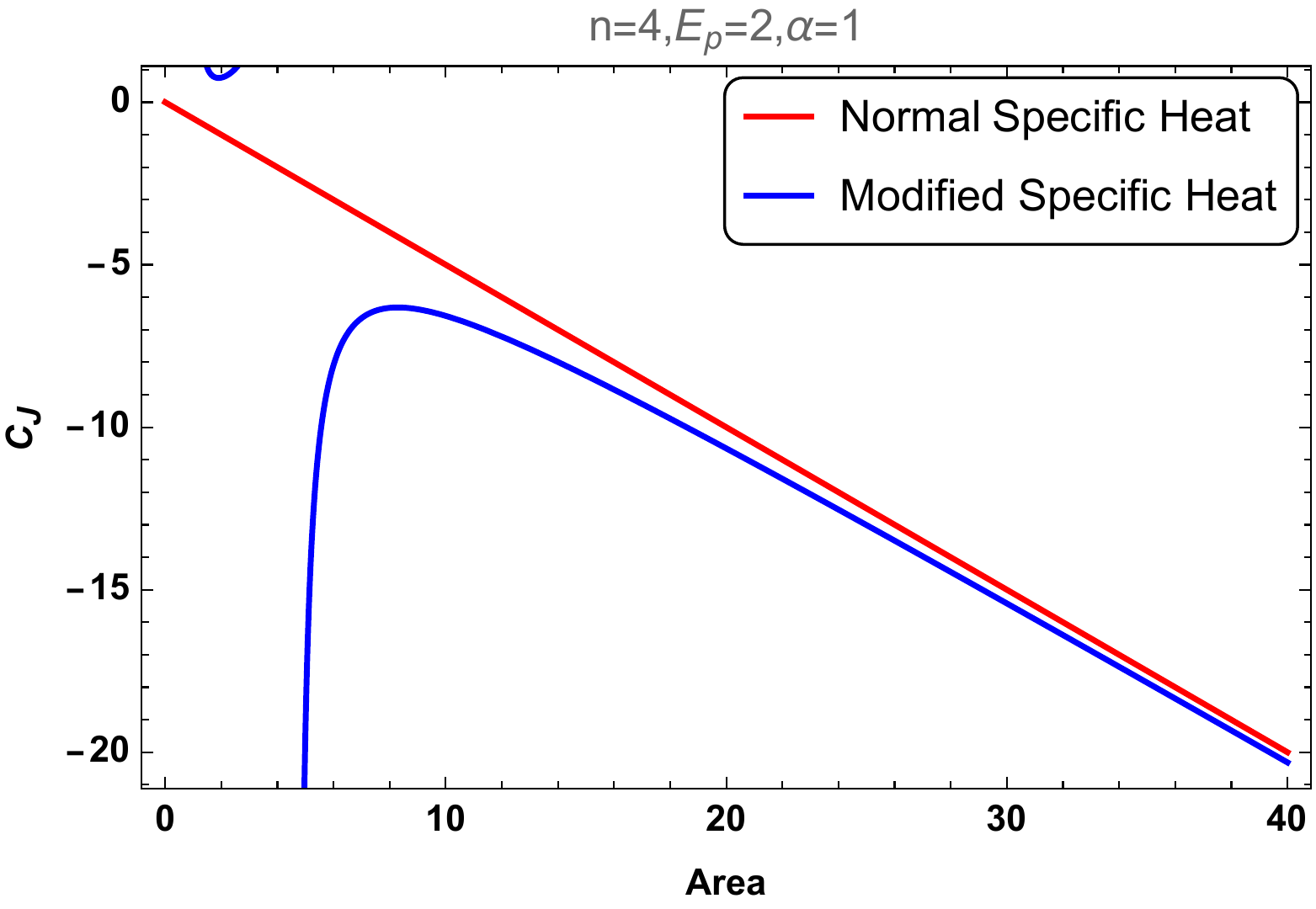}
    \caption{Specific heat $C_J$ versus area of Schwarzschild black hole in  gravity's rainbow and normal gravity}
    \label{fig:scCA}
\end{figure}

\begin{figure}[H]
    \centering
    \includegraphics[scale=0.7]{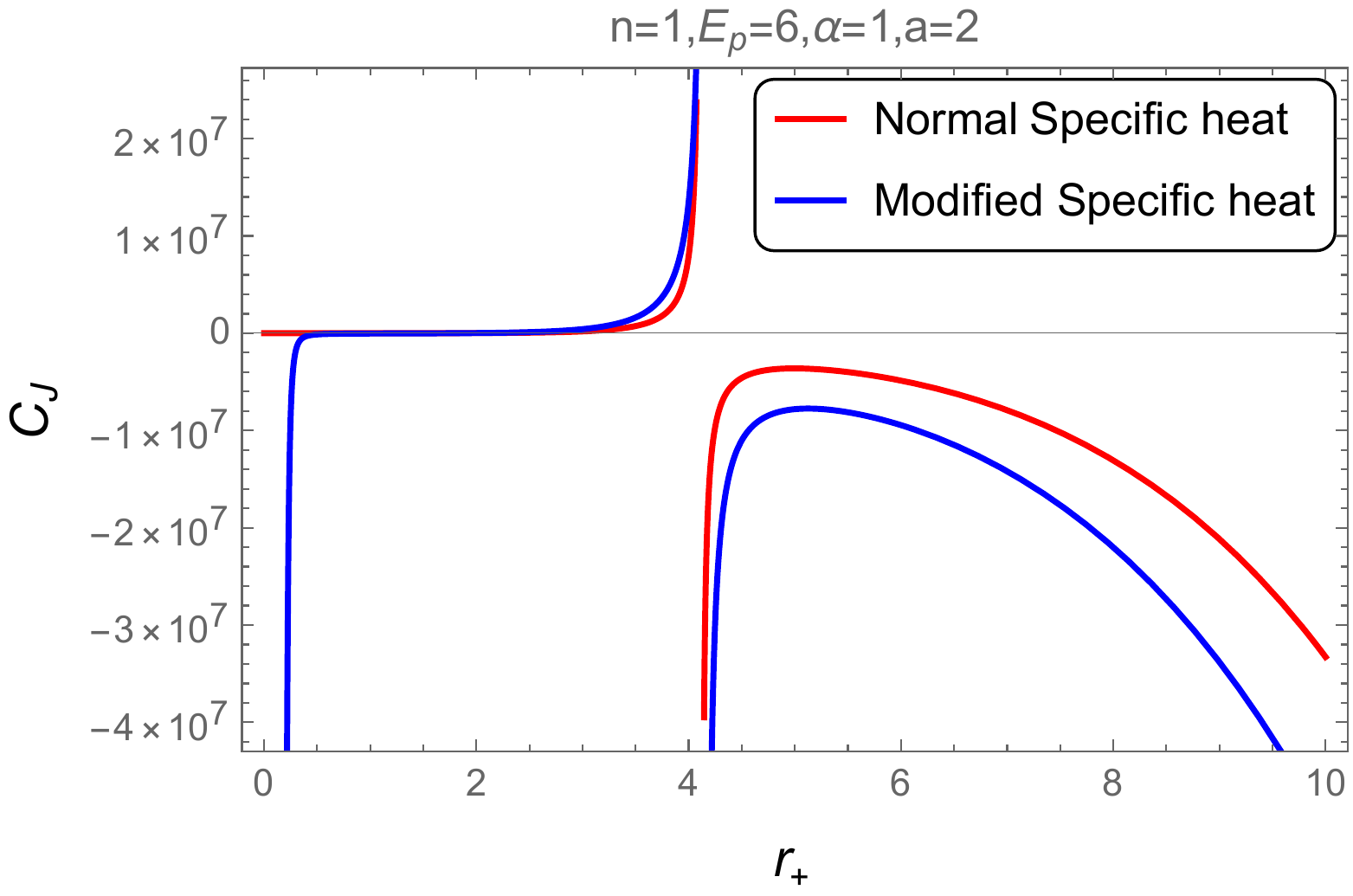}
    \caption{Normal and modified Specific heat of Kerr black hole}
    \label{fig:kCr}
\end{figure}

Now we will analyze a Kerr black hole in  gravity's rainbow. We will again use the    rainbow function motivated from    the hard spectra from gamma-ray bursters at cosmological distances \cite{amelino1998tests}.
Rotating black holes in flat and (A)dS space can be cast in the form\cite{ma2008hawking},
\begin{align}
    {ds}^2=-A(r,\theta ){dt}^2+\frac{1}{B(r,\theta )}{dr}^2+g_{\theta \theta }\;{d\theta }^2+g_{\phi \phi \;}{d\phi }^2-2g_{{t\phi
}}\;{dt} {d\phi }
\end{align}
The temperature for this metric is given by\cite{wald2010general,ma2008hawking},
\begin{align}
   T=\frac{1}{4\pi }\sqrt{A_{,r}\big(r_+,0\big) B_{,r}\big(r_+,0\big)}
\end{align}
Again $r_+$ is given by $B(r_+)=0$. The metric for Kerr black hole is given by\cite{altamirano2014thermodynamics}
\begin{align}
    {ds}^2=-{dt}^2+\frac{2M r}{\Sigma }\big({dt}-a\; {Sin}^2\theta  {d\phi }\big)^2+\frac{\Sigma }{\Delta }{dr}^2+\Sigma
    {d\theta }^2+\big(r^2+a^2\big){Sin}^2\theta  {d\phi }^2
\end{align}
where $\Sigma =r^2+a^2{Cos}^2\theta $ and $\Delta =r^2+a^2-2M r$.
The temperature for Kerr black hole is given by\cite{altamirano2014thermodynamics,wald2010general}
\begin{align}
    T_0=\frac{1}{2\pi }\bigg(\frac{r_+}{a^2+r_+{}^2}-\frac{1}{2r_+}\bigg)
\end{align}
which modifies in  gravity's rainbow to  be
\begin{align}
    T=\frac{1}{2\pi }\bigg(\frac{r_+}{a^2+r_+{}^2}-\frac{1}{2r_+}\bigg)\frac{\alpha \big(\frac{1}{r_+E_p}\big){}^n}{e^{\alpha \big(\frac{1}{r_+E_p}\big){}^n}-1}
\end{align}
To get the modified entropy from modified temperature we use the first law of thermodynamics,
\begin{align}
    {{dS}=\frac{{dM}}{T}-\frac{\Omega }{T}{dJ}-\frac{\Phi }{T}{dQ}}
\end{align}
For Kerr black hole,
\begin{align}
    \Omega =\frac{a}{a^2+r_+{}^2}, \hspace{2cm}  J=\frac{a\big(a^2+r_+{}^2\big)}{2r_+},   \hspace{1cm} Q=0
\end{align}
After simplification, the modified entropy turns out to be,
\begin{align}
    S=\frac{2 E_p{}^n \pi }{\alpha } \int r_+{}^{n+1}\big(e^{\alpha \big(\frac{1}{r_+E_p}\big){}^n}-1\big)dr_+
\end{align}
The thermodynamic stability of black holes is determined by the heat capacity $C_J$ at constant angular momentum $J$. Using the modified temperature and entropy in equation \eqref{cj}, we plot the specific heat of Kerr black hole in figure \eqref{fig:kCr},
\begin{figure}[H]
    \centering
    \includegraphics[scale=0.7]{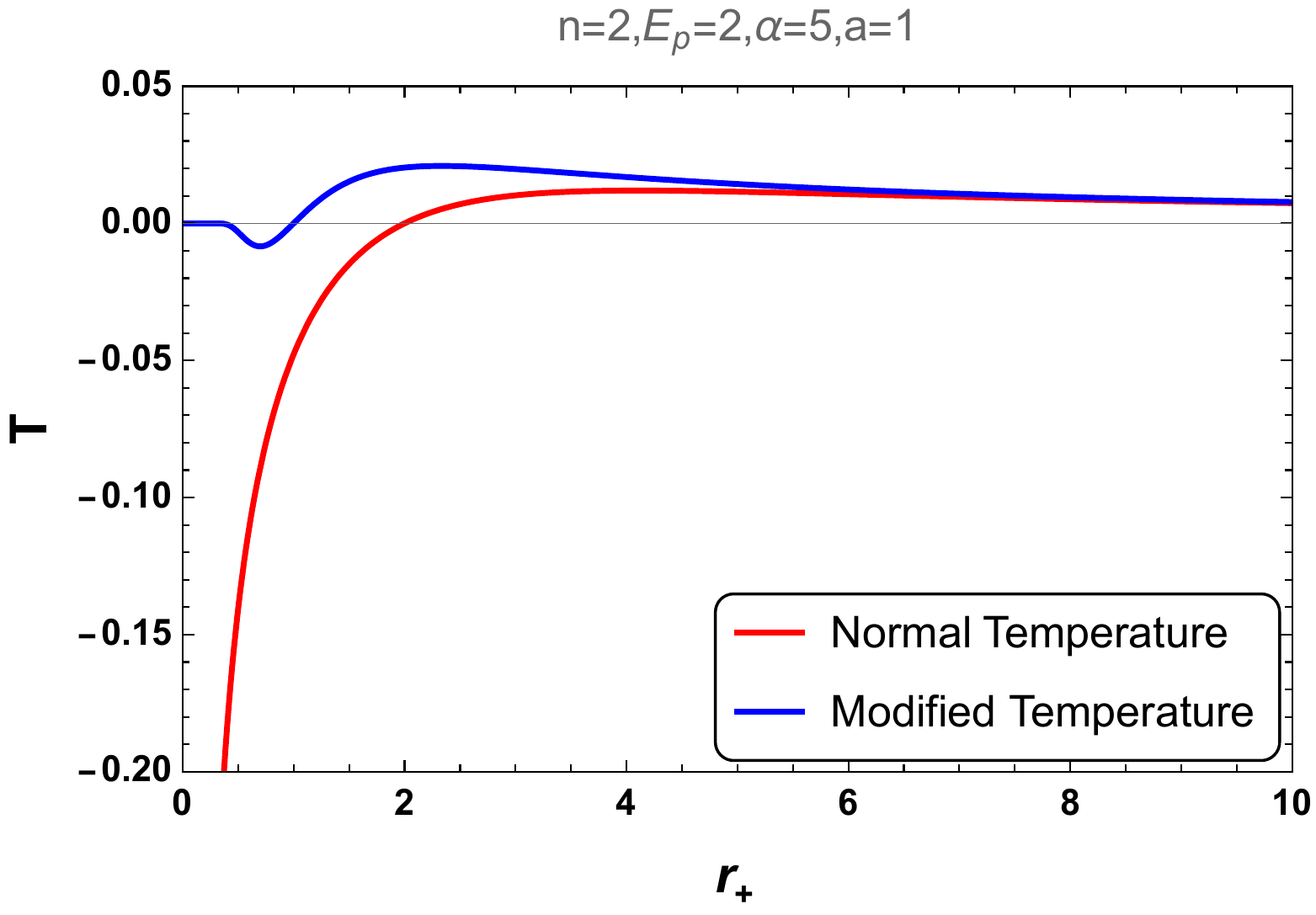}
    \caption{Temperature of Kerr black hole in normal gravity and  gravity's rainbow versus $r_+$}
    \label{fig:kTr}
\end{figure}

\begin{figure}[H]
    \centering
    \includegraphics[scale=0.7]{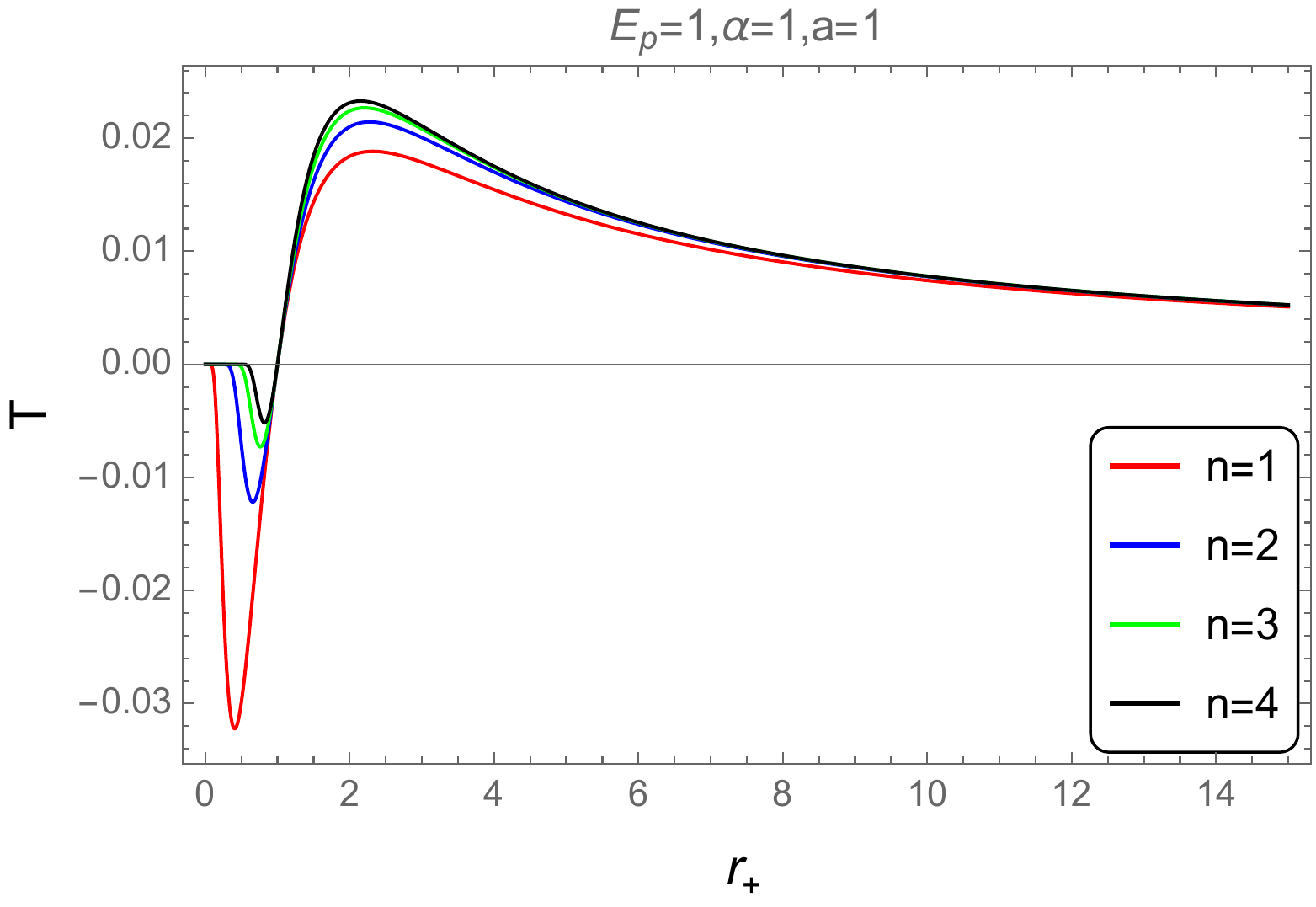}
    \caption{Comparison of Kerr black hole's temperature in  gravity's rainbow for different values of $n=1,2,3,4$.}
    \label{fig:kTrR}
\end{figure}

\begin{figure}[H]
    \centering
    \includegraphics[scale=0.7]{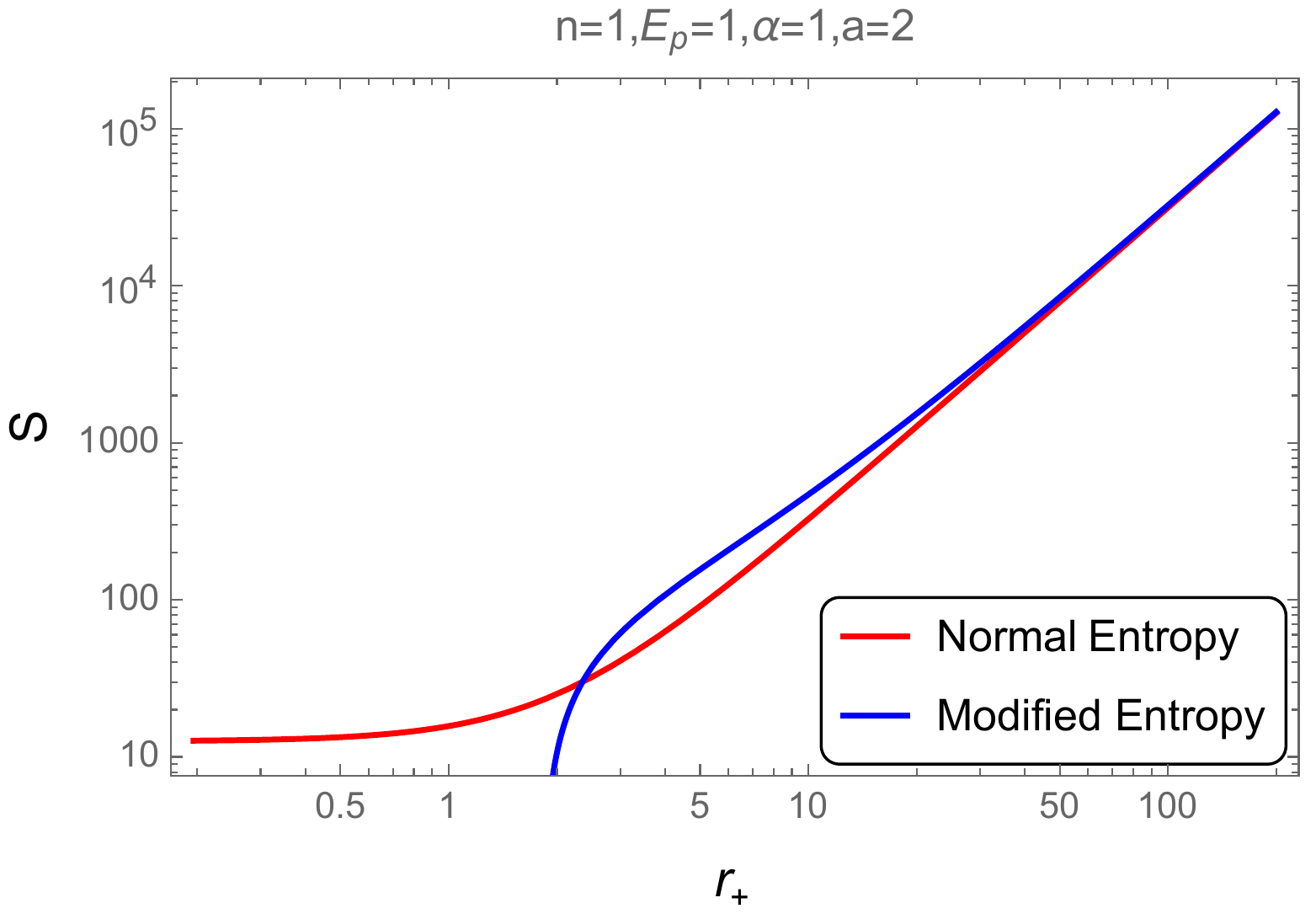}
    \caption{Entropy of Kerr black hole in normal and  gravity's rainbow}
    \label{fig:kSr}
\end{figure}

It was observed that at large scales, the modification due to rainbow functions can be neglected. However, at small scales, this modification produces important changes. 
Due to the modification by rainbow functions, both temperature, and entropy for these black holes reduces to zero at finite radius. As the temperature reduces to zero these black holes do not radiate and will not reduce in size. Thus, they will form black black remnants. The formation of black remnants due to rainbow functions from loop quantum gravity is well known \cite{Amelino-Camelia:2008aez}. However, here we have obtained this from rainbow functions motivated from hard spectra from gamma-ray bursters at cosmological distances. Thus, these results hold for astrophysical black holes.  Furthermore, at this radius, the specific heat also vanishes, and this demonstrates that the black holes do not radiate with the surroundings. These will form the last stage of evaporation of black holes. It has been proposed that  the virtual black holes occur at the stages of evaporation of black holes \cite{hawking1996virtual,hawking1997loss,faizal2012some,ohkuwa2017virtual}. Thus,  these black remnants will  form stable virtual black holes  in to gravity's rainbow.  These  virtual black holes can be used to resolve information paradox \cite{hawking1996virtual,hawking1997loss,faizal2012some,ohkuwa2017virtual}. Now we have observed that they will be stable due to rainbow functions, and this can resolve the information paradox. As the information can reside in these stable virtual black holes at the end of the evaporation of black holes. Unlike the usual picture, where a pair of black holes form and annihilate, here a form of virtual black holes form and do not annihilate. This is similar to what happens for particles in Hawking radiation. It would be interesting to investigate  the consequences of this model, and if it can be observed using any cosmological or astrophysical observation.  So, in this letter, we analyzed the effect of rainbow function on the thermodynamics of black holes. 
This was done for Schwarzschild  black hole and Kerr black holes. We used rainbow function motivated from the hard spectra from gamma-ray bursters at cosmological distances. We obtained black remnants due to this modification, and proposed them to be stable virtual black holes. 
\section{Acknowledgement}
M. V. Akram would like to thank Prof. S. Barve for useful discussions. 
\bibliographystyle{unsrt}
\bibliography{bibl.bib}

\begin{thebibliography}{10}

\bibitem{rosten2012fundamentals}
OJ~Rosten.
\newblock Fundamentals of the exact renormalization group, phys, 2012.

\bibitem{warner2000renormalization}
Nicholas~P Warner.
\newblock Renormalization-group flows from five-dimensional supergravity.
\newblock {\em Classical and Quantum Gravity}, 17(5):1287, 2000.

\bibitem{de2018gravitational}
Claudia de~Rham and Scott Melville.
\newblock Gravitational rainbows: Ligo and dark energy at its cutoff.
\newblock {\em Physical review letters}, 121(22):221101, 2018.

\bibitem{ali2015proposal}
Ahmed~Farag Ali and Mohammed~M Khalil.
\newblock A proposal for testing gravity's rainbow.
\newblock {\em EPL (Europhysics Letters)}, 110(2):20009, 2015.

\bibitem{assanioussi2017rainbow}
Mehdi Assanioussi and Andrea Dapor.
\newblock Rainbow metric from quantum gravity: Anisotropic cosmology.
\newblock {\em Physical Review D}, 95(6):063513, 2017.

\bibitem{hendi2015black}
Seyed~Hossein Hendi and Mir Faizal.
\newblock Black holes in gauss-bonnet gravity’s rainbow.
\newblock {\em Physical Review D}, 92(4):044027, 2015.

\bibitem{amelino1997distance}
G~Amelino-Camelia, John Ellis, NE~Mavromatos, and Dimitri~V Nanopoulos.
\newblock Distance measurement and wave dispersion in a liouville-string
  approach to quantum gravity.
\newblock {\em International Journal of Modern Physics A}, 12(03):607--623,
  1997.

\bibitem{amelino2013quantum}
Giovanni Amelino-Camelia.
\newblock Quantum-spacetime phenomenology.
\newblock {\em Living Reviews in Relativity}, 16(1):1--137, 2013.

\bibitem{hooft1996quantization}
G't Hooft.
\newblock Quantization of point particles in (2+ 1)-dimensional gravity and
  spacetime discreteness.
\newblock {\em Classical and Quantum Gravity}, 13(5):1023, 1996.

\bibitem{kostelecky1989spontaneous}
V~Alan Kosteleck{\`y} and Stuart Samuel.
\newblock Spontaneous breaking of lorentz symmetry in string theory.
\newblock {\em Physical Review D}, 39(2):683, 1989.

\bibitem{gambini1999nonstandard}
Rodolfo Gambini and Jorge Pullin.
\newblock Nonstandard optics from quantum space-time.
\newblock {\em Physical Review D}, 59(12):124021, 1999.

\bibitem{carroll2001noncommutative}
Sean~M Carroll, Jeffrey~A Harvey, V~Alan Kosteleck{\`y}, Charles~D Lane, and
  Takemi Okamoto.
\newblock Noncommutative field theory and lorentz violation.
\newblock {\em Physical Review Letters}, 87(14):141601, 2001.

\bibitem{faizal2013noncommutative}
Mir Faizal.
\newblock Noncommutative quantum gravity.
\newblock {\em Modern Physics Letters A}, 28(10):1350034, 2013.

\bibitem{faizal2011spontaneous}
Mir Faizal.
\newblock Spontaneous breaking of lorentz symmetry by ghost condensation in
  perturbative quantum gravity.
\newblock {\em Journal of Physics A: Mathematical and Theoretical},
  44(40):402001, 2011.

\bibitem{ali2014remnants}
Ahmed~Farag Ali, Mir Faizal, and Mohammed~M Khalil.
\newblock Remnants of black rings from gravity’s rainbow.
\newblock {\em Journal of High Energy Physics}, 2014(12):1--14, 2014.

\bibitem{ali2015remnant}
Ahmed~Farag Ali, Mir Faizal, and Mohammed~M Khalil.
\newblock Remnant for all black objects due to gravity's rainbow.
\newblock {\em Nuclear Physics B}, 894:341--360, 2015.

\bibitem{ali2015absence}
Ahmed~Farag Ali, Mir Faizal, and Mohammed~M Khalil.
\newblock Absence of black holes at lhc due to gravity's rainbow.
\newblock {\em Physics Letters B}, 743:295--300, 2015.

\bibitem{ashour2016branes}
Amani Ashour, Mir Faizal, Ahmed~Farag Ali, and Fay{\c{c}}al Hammad.
\newblock Branes in gravity’s rainbow.
\newblock {\em The European Physical Journal C}, 76(5):1--9, 2016.

\bibitem{heydarzade2017vaidya}
Yaghoub Heydarzade, Prabir Rudra, Farhad Darabi, Ahmed~Farag Ali, and Mir
  Faizal.
\newblock Vaidya spacetime in massive gravity's rainbow.
\newblock {\em Physics Letters B}, 774:46--53, 2017.

\bibitem{Rudra:2016alu}
Prabir Rudra, Mir Faizal, and Ahmed~Farag Ali.
\newblock {Vaidya Spacetime for Galileon Gravity's Rainbow}.
\newblock {\em Nucl. Phys. B}, 909:725--736, 2016.

\bibitem{Aounallah:2020yjo}
Houcine Aounallah, Behnam Pourhassan, Seyed~Hossein Hendi, and Mir Faizal.
\newblock {Five-dimensional Yang\textendash{}Mills black holes in massive
  gravity\textquoteright{}s rainbow}.
\newblock {\em Eur. Phys. J. C}, 82(4):351, 2022.

\bibitem{Hendi:2017pld}
Seyed~Hossein Hendi, Ali Dehghani, and Mir Faizal.
\newblock {Black hole thermodynamics in Lovelock gravity's rainbow with (A)dS
  asymptote}.
\newblock {\em Nucl. Phys. B}, 914:117--137, 2017.

\bibitem{Hendi:2016njy}
Seyed~Hossein Hendi, Shahram Panahiyan, Behzad Eslam~Panah, Mir Faizal, and
  Mehrab Momennia.
\newblock {Critical behavior of charged black holes in Gauss-Bonnet
  gravity\textquoteright{}s rainbow}.
\newblock {\em Phys. Rev. D}, 94(2):024028, 2016.

\bibitem{Hendi:2015hja}
Seyed~Hossein Hendi and Mir Faizal.
\newblock {Black holes in Gauss-Bonnet gravity\textquoteright{}s rainbow}.
\newblock {\em Phys. Rev. D}, 92(4):044027, 2015.

\bibitem{Hendi:2015cra}
S.~H. Hendi, Mir Faizal, B.~Eslam Panah, and S.~Panahiyan.
\newblock {Charged dilatonic black holes in gravity\textquoteright{}s rainbow}.
\newblock {\em Eur. Phys. J. C}, 76(5):296, 2016.

\bibitem{Amelino-Camelia:2008aez}
Giovanni Amelino-Camelia.
\newblock {Quantum-Spacetime Phenomenology}.
\newblock {\em Living Rev. Rel.}, 16:5, 2013.

\bibitem{amelino1998tests}
Giovanni Amelino-Camelia, John Ellis, NE~Mavromatos, Dimitri~V Nanopoulos, and
  Subir Sarkar.
\newblock Tests of quantum gravity from observations of $\gamma$-ray bursts.
\newblock {\em Nature}, 393(6687):763--765, 1998.

\bibitem{hawking1996virtual}
Stephen~W Hawking.
\newblock Virtual black holes.
\newblock {\em Physical Review D}, 53(6):3099, 1996.

\bibitem{hawking1997loss}
Stephen~William Hawking and Simon~F Ross.
\newblock Loss of quantum coherence through scattering off virtual black holes.
\newblock {\em Physical Review D}, 56(10):6403, 1997.

\bibitem{faizal2012some}
Mir Faizal.
\newblock Some aspects of virtual black holes.
\newblock {\em Journal of Experimental and Theoretical Physics},
  114(3):400--405, 2012.

\bibitem{ohkuwa2017virtual}
Yoshiaki Ohkuwa, Mir Faizal, and Yasuo Ezawa.
\newblock Virtual black holes in a third quantized formalism.
\newblock {\em Annals of Physics}, 384:105--115, 2017.

\bibitem{Ali:2015iba}
Ahmed~Farag Ali, Mir Faizal, Barun Majumder, and Ravi Mistry.
\newblock {Gravitational Collapse in Gravity's Rainbow}.
\newblock {\em Int. J. Geom. Meth. Mod. Phys.}, 12(09):1550085, 2015.

\bibitem{Ali:2014cpa}
Ahmed~Farag Ali, Mir Faizal, and Barun Majumder.
\newblock {Absence of an Effective Horizon for Black Holes in Gravity's
  Rainbow}.
\newblock {\em EPL}, 109(2):20001, 2015.

\bibitem{angheben2005hawking}
Marco Angheben, Mario Nadalini, Luciano Vanzo, and Sergio Zerbini.
\newblock Hawking radiation as tunneling for extremal and rotating black holes.
\newblock {\em Journal of High Energy Physics}, 2005(05):014, 2005.

\bibitem{ali2014black}
Ahmed~Farag Ali.
\newblock Black hole remnant from gravity’s rainbow.
\newblock {\em Physical Review D}, 89(10):104040, 2014.

\bibitem{ma2008hawking}
Zheng~Ze Ma.
\newblock Hawking temperature of kerr--newman--ads black hole from tunneling.
\newblock {\em Physics letters B}, 666(4):376--381, 2008.

\bibitem{wald2010general}
Robert~M Wald.
\newblock {\em General relativity}.
\newblock University of Chicago press, 2010.

\bibitem{altamirano2014thermodynamics}
Natacha Altamirano, David Kubiz{\v{n}}{\'a}k, Robert~B Mann, and Zeinab
  Sherkatghanad.
\newblock Thermodynamics of rotating black holes and black rings: phase
  transitions and thermodynamic volume.
\newblock {\em Galaxies}, 2(1):89--159, 2014.

\end{thebibliography}
\end{document}